\documentstyle[aps,eqsecnum,twoside]{revtex}
\newcommand{\C}{{\if mm {{\rm C}\mkern -15mu{\phantom{\rm t}\vrule}}
\mkern +10mu \else \leavemode \hbox{I}\kern -.17em \hbox{C} \fi}}
\newcommand {\bJ}{\mbox{\bf J}} 
 
\newcommand {\bj}{\mbox{\bf j}} 
\newcommand {\br}{\mbox{\bf r}}

\newcommand {\bv}{\mbox{\bf v}}
\newcommand {\bA}{\mbox{\bf A}}
\newcommand {\bB}{\mbox{\bf B}}
\newcommand {\bP}{\mbox{\bf P}}
\newcommand {\bM}{\mbox{\bf M}}

\newcommand {\bD}{\mbox{\bf D}}
\newcommand {\bH}{\mbox{\bf H}}
\newcommand {\bE}{\mbox{\bf E}}
\newcommand {\bq}{\mbox{\bf q}}

\newcommand {\bpi}{\mbox{\boldmath $\Pi$}}

\newcommand {\bt}{\mbox{\bf t}}
\newcommand {\bT}{\mbox{\bf T}}
\newcommand {\bal}{\mbox{\boldmath $\alpha$}}
\newcommand {\bbe}{\mbox{\boldmath $\beta$}}

\newcommand {\cE}{\mbox{\boldmath $\cal E$}}
\newcommand {\cB}{\mbox{\boldmath $\cal B$}}
\newcommand {\cD}{\mbox{\boldmath $\cal D$}}
\newcommand {\cH}{\mbox{\boldmath $\cal H$}}
\newcommand {\bp}{\mbox{\bf p}}
\newcommand {\p}{\partial}

\newcommand {\bC}{\mbox{\bf C}}
\newcommand {\bV}{\mbox{\bf V}}

\newcommand {\bR}{\mbox{\bf R}}
\newcommand {\pr}{\prime}
\newcommand {\vf}{\varphi}

\begin{document}
\title{Electrodymanics of Moving Continuous Media with Toroid Polarization}
\author{V.M.~Dubovik and B.~Saha \\
{\small \it Bogoliubov Lab. of Theoretical Physics }\\
{\small \it Joint Institute for Nuclear Research\\
141980, Dubna, Moscow reg., Russia\\
e-mail: dubovik@thsun1.jinr.dubna.su\\  
e-mail: saha@thsun1.jinr.dubna.su}}
\maketitle
\begin{abstract}

With regard to the toroid contributions, a modified system of 
equations of electrodynamics moving continuous media has been obtained.
Alternative formalisms to introduce the toroid moment contributions 
in the equations of electromagnetism has been worked out.
The two four-potential formalism has been developed for the electromagnetic
continous media subjected to Lorentz transformations.
 
\end{abstract}
 Maxwell theory: general mathematical aspects
\vskip 5mm
\section{Introduction} 
\setcounter{equation}{0} 
The history of electromagnetism is the history of struggle of different
rival concepts from the very early days of its existence. Though, after
the historical observation by Hertz, all main investigations in
electromagnetism were based on Maxwell equations, nevertheless this
theory still suffers from some shortcomings inherent to its predecessors.
Several attempts were made to remove the internal inconsistencies of the 
theory. To be short we refer to very few of them. One of the attempts to
modify the theory of electromanetism was connected with the introduction
of magnetic charge in Maxwell equation by Dirac~\cite{Dirac,Dirac1}, 
while keeping the usual definition of $\bE$ and $\bB$ in terms of the gauge 
potentials. A very interesting work in this direction was done by Miller
~\cite{MillerRF}. In this paper he showed the mutual substitution of the
sources as follows: 
$$\rho^e \rightarrow \rho^e \bv = \bj^e = c {\rm curl}\bM \rightarrow
 -{\rm div} \bM = \rho^m \rightarrow \rho^m \bv = j^m = c {\rm curl} \bP
\rightarrow -{\rm div} \bP = \rho^e \rightarrow$$
Recently D. Singleton~\cite{Singleton,Singleton1} gave an alternative
formulation of classical electromagnetism with magnetic and electric charges
by introducing two four-vector potentials 
$A^\mu = (\phi_e, \bA)$ and $C^\mu = (\phi_m, \bC)$ and defining 
$\bE$ and $\bB$ fields as
\begin{mathletters}
\begin{eqnarray}
\bE &=& - \nabla \phi_e - \frac{\p \bA}{\p t} - \mbox{curl}\,\bC \\
\bB &=& - \nabla \phi_m - \frac{\p \bC}{\p t} + \mbox{curl}\,\bA
\end{eqnarray}
\end{mathletters}
Inserting these newly defined vector potentials into the generalized 
Maxwell equations~\cite{Jackson}
\begin{mathletters}
\begin{eqnarray}
{\rm curl}\, \bB &=& \frac{\p \bE}{\p t} + \bJ_e  \\  
{\rm div}\, \bE &=&  \rho_e  \\
- {\rm curl}\, \bE &=& \frac{\p \bB}{\p t} + \bJ_m  \\
{\rm div}\, \bB &=& \rho_m  
\end{eqnarray}
\end{mathletters}
where $\rho_m$ and $\bJ_m$ are the magnetic charge and current, respectively,
and imposing the Lorentz gauge condition
\begin{equation}
\frac{\p \phi_e}{\p t} + {\rm div}\,\bA = 0, \quad
\frac{\p \phi_m}{\p t} + {\rm div}\,\bC = 0
\end{equation}
he arrived at the wave form of Maxwell's equations with electric and
magnetic charges. Note that, a similar theory (two potential formalism) was
developed by us few years ago (we will come back to it in Sec. 3). 
Though this treatment avoids the use of singular, nonlocal variables
in electrodynamics with magnetic charge and makes the Maxwell system
more symmetric, since both the charges in this approach are gauge charges, 
the main defect of this theory in our view is that
the existence of magnetic charge still lack of experimental support,
hence can be considered as a mathematically convenient one only.

Recently Chubykalo a.o. made an effort to modify the electromagnetic 
theory by invoking both the transverse and longitudinal (explicitly 
time independent) fields simultaneously, thus giving an equal footing 
to both the Maxwell-Hertz and Maxwell-Lorentz equations
~\cite{Andrew,Chubykalo,Evans}. To remove all ambiguities related to the 
applications of Maxwell's displacement current they substituted all partial 
derivatives in Maxwell-Lorenz equations by the {\it total} ones, such that
\begin{equation}
\frac{d}{d t} = \frac{\p}{\p t} - (\bV \cdot {\rm grad}) \label{in1}
\end{equation} 
where $\bV = d \br_q/dt$ and $\br_q (t)$ are the velocity and the coordinate 
of the charge, respectively, at the instant $t$. Further they separated
all field quantities into {\it two independent classes} with explicit 
$\{\}^*$ and implicit $\{\}_0$ time dependence, respectively. Thus, the
component $\bE_0$ of the total electric field $\bE$ in every point is
understood to depend {\it only} on the position of source at a given instant.
In other words, $\bE_0$ is rigidly linked with the location of the charge.
From this point of view, the partial time derivative in~(\ref{in1}) must be
related {\it only} with the explicit time-dependent component $\bE^*$ 
whereas the convection derivative {\it only} with $\bE_0$:
\begin{equation}
\frac{d\bE}{d t} = \frac{\p \bE^*}{\p t} - (\bV \cdot {\rm grad})\bE_0, \quad
\bE = \bE^* (\br, t) + \bE_0 (\bR (t))
\end{equation} 
where $\br$ is a fixed distance from the origin of the reference system at
rest to the point of observation and $\bR (t) = \br - \br_q (t)$.

Another attempt to modify the equations of eletromagnetism is connected
with the existence of the third family of multipole moments, namely
the {\it toroid} one. This theory was developed by us during the recent
years. Recently we introduced toroid moments in Maxwell equations exploiting
Lagrangian formalism~\cite{ICTP}. In the Sec. 2 of this paper we give a brief
description of this formalism. Moreover, here we develop an alternative
method to introduce toroid moments in the equation of electromagnetism.
In Sec. 3 we develop two potential formalism suggested by us earlier. 

\section{Introduction of toroid moments in the equations of electromagnetism} 
\setcounter{equation}{0}
\vskip 3mm
In early fifties, while solving the problem of the multipole radiation 
of a spatially
bounded source, Franz and Wallace~\cite{Franz,Wallace} found a contribution to
the electric part of radiation at the expense of magnetization. Further
Ya.~Zel'dovich~\cite{Zel'dovich} pointed out the non-correspondence between
the existence of two known multipole sets, Coulomb and magnetic, and the
number of form-factors for a spin -- $\frac{1}{2}$ charged particles. 
Following the parity non-conservation law in weak interactions Zel'dovich
suggested a third form-factor in the parametrization of the Dirac spinor
particle current. As a classical counterpart of this form-factor he 
introduced {\it anapole} in connection with the global electromagnetic
properties of a toroid coil that are impossible to describe within the
charge or magnetic dipole moments in spite of explicit axial symmetry
of the toroid coil. In 1965 Shirokov and Cheshkov~\cite{Shirokov} constructed
the parametrization for relativistic matrix elements of currents of 
charged and spinning particle, which contain
the third set of form-factors. Finally, in 1974 Dubovik and Cheskov~\cite{DC} 
determined the toroid moment in the framework of classical electrodynamics.
Note that {\it anapole} and {\it toroid dipole} are not the different
names of one and the same thing. They are indeed quite differnt in nature.
For example, the anapole cannot radiate at all while the toroid coil 
and its point-like model, toroid dipole, can. The 
matter is that the anapole is some composition of electric dipole and actual 
toroid dipole giving destructive interference of their radiation.
Recently a principally new type of magnetism known as {\it aromagnetism} was 
observed in a class of organic substances, suspended either in water or in 
other liquids~\cite{Spartakov}. Later, it was shown that this phenomena
of aromagnetism cannot be explained in a standard way, e.g., by 
ferromegnetism, since the organic molecules do not possess magnetic moments
of either orbital or spin origin. It was also shown that the origin of
aromagnetism is the interaction of vortex electric field induced by 
alternative magnetic one with the axial toroid moments in aromatic substances
~\cite{MM}. 

In a recent work Dubovik and Kuznetsov~\cite{Kuznetsov} calculated the toroid
moment of Majorana neutrino. It was also pointed out that the magnitude of
the toroid dipole moment of a Dirac neutrino ($\nu_d$) is just the half of 
that of a majorana one ($\nu_m$) and both of them posesses non-trivial
torid moments even if $m_\nu = 0 ~\cite{Bukina}.$

The latest theoretical and experimental development force the introduction 
of toroid moments in the framework of conventional classical electrodynamics 
that in its part inevitably leads to the modification of the equations
of electromagnetism and the equations of motion of particles in external
electromagnetic field. In the two following subsection we give two
alternative schemes of introduction of toroid polarizations in the 
electromagnetic equations.

To begin with we write the Maxwell equations for electromagnetic fields in 
vacuum, in the presence of extraneous electric charge $\rho$ and electric 
current, that is, charge - in - motion, of density $\bj$. 
\begin{mathletters}
\label{Eq:M}
\begin{eqnarray}
{\rm curl} \bB - \frac{1}{c}\frac{\p \bE}{\p t} &=& 
\frac{4 \pi}{c} \bj  \label{Ma}  \\  
{\rm div} \bE &=& 4 \pi \rho  \label{Mb}\\
{\rm curl} \bE + \frac{1}{c}\frac{\p \bB}{\p t} &=& 0 \label{Mc}  \\
{\rm div} \bB &=& 0  \label{Md} 
\end{eqnarray}
\end{mathletters}
where $\bE$ and $\bB$ are the flux densities of electric field and 
magnetic induction, respectively. Note that, the electric charges and
electric currents, being destributed in vacuum, construct the 
electromagnetic structure of matter~\cite{Turov} and are relateted to the 
elementary charge $e$ in the following way
\begin{mathletters}
\label{MCC}
\begin{eqnarray}
\rho (\br) &=& \sum_{n} e_n \delta (\br - \bq_n) \label{charge}\\
\bj (\br) &=& \sum_{n} e_n \dot{\bq}_n \delta (\br - \bq_n) \label{current}
\end{eqnarray}
\end{mathletters}
So, to describe the system as a whole the equations~(\ref{Eq:M}) and 
~(\ref{MCC}) should be supllemented by the equation of motion of 
micro-particle, i.e.,
\begin{equation}
\label{LF} 
m_n \ddot{\bq}_n = e_n {\bE} + \frac{e_n}{c}\dot{\bq}_n \times {\bB}
\end{equation}
In this process there occurs a vast number of problems connected with
the different areas of this vast field:\\
Write the classical and quantum analogies of the equations of motion
of a point-like particle possessing toroid dipole (with usual properties);\\
Solve the boundary-value problems for the model with all the of vector
polarizations;\\
Formulate the electrodynamics of continuous infinite media for the latter
case. 

Since the electromagnetic field in media generates bound charges and
bound currents, the source parts in~(\ref{Ma}) and ~(\ref{Mb}) should be 
supplemented as follows~\cite{Purcell} 
\begin{mathletters}
\label{bound}
\begin{eqnarray}
\rho_{\rm total} &=& \rho + \rho_{\rm bound} = \rho - 
{\rm div} \bP  \label{b1}\\
\bj_{\rm total} &=& \bJ = \bj + \frac{\p \bP}{\p t} 
+ c\, {\rm curl}\, \bM \label{b2}
\end{eqnarray}
\end{mathletters}
Then we may write the Maxwell equations for electromagnetic fields in media 
as  
\begin{mathletters}
\label{Eq:MM}
\begin{eqnarray}
{\rm curl} \bH - \frac{1}{c}\frac{\p \bD}{\p t} &=& 
\frac{4 \pi}{c} \bj  \label{MMa}  \\  
{\rm div} \bD &=& 4 \pi \rho  \label{MMb}\\
{\rm curl} \bE + \frac{1}{c}\frac{\p \bB}{\p t} &=& 0 \label{MMc}  \\
{\rm div} \bB &=& 0  \label{MMd} 
\end{eqnarray}
\end{mathletters}
where we denote 
$\bD = \bE + 4 \pi \bP$ and $\bH = \bB - 4 \pi \bM$ are the flux densities
of electric displacement vector and magnetic field, respectively.
 
Note that $\p P /\p t$, known as bound-charge current density is independent
of the details of the model and is a conduction current. The difference 
between an ``ordinary'' conduction current density and the current density
$\p P /\p t$ is that the first involves {\it free} charge $\rho$ (foreign
charge over which we have some control, i.e., charge that can be added to
or remove from an object) in motion, when the second {\it bound} charge 
(the integral parts of atoms or molecules of the dielectric) in motion. 
Another obvious practial distinction is that it is impossible to get a 
{\it steady} bound charge current that goes forever unchanged. The 
{\it bound} currents $c\,{\rm curl} \bM$ are associated with molecular or 
atomic magnetic moments, including the intrinsic magnetic moment of 
particles with spin, whereas {\it free} currents  $\bj$ are ``ordinary'' 
conduction currents flowing on macroscopic paths and can be started or 
stopped with a switch and measured with an ammeter.   

The foregoing equations can be obtained from a Lagrangian describing
the interacting system of electromagnetic field and non-relativistic 
test charged particle~\cite{Healy}
\begin{eqnarray} 
\label{L1}
L =  \frac{1}{2} m \dot{\bq}^{2} 
+  \frac{1}{8\pi} \int \bigl[\frac{\dot{\bA}      
^2}{c^2} - (\mbox{curl} {\bA})^2\bigr]  d\br 
+ \frac{1}{c} \int {\bJ}({\br}) \cdot {\bA}({\br}) d\br 
\end{eqnarray}
Here the total current of the system
$\bJ$ coincides with the current $\bj = e \dot{q}$, generated by the free 
charges (charged particles) in motion since there is no polarization current. 

It can be easily verified that variation of the Lagrangian~(\ref{L1})
with respect to the particle coordinates gives the second law of Newton 
with the Lorentz force, i.e., the equation ~(\ref{LF})  
\begin{eqnarray}
m\,\ddot{\bq}\,=\,e\,{\bE}({\bq},\,t)\,+\,
\frac{e}{c}\dot{\bq} \times {\bB}({\bq},\,t) \nonumber
\end{eqnarray}
whreas the variation with respect to field variables gives~(\ref{Ma}), i.e.,
\begin{eqnarray}
{\rm curl} \bB - \frac{1}{c}\frac{\p \bE}{\p t}
= -\frac{4\pi}{c} {\bj} \label{CE}
\end{eqnarray}
whre we denote 
$\bB = {\rm curl} \bA$ and $\bE = - c^{-1} (\p \bA /\p t)$.
It should be emphasized that $\bE$ in ~(\ref{LF}) and ~(\ref{CE}) is
the transverse part of the total electric field. The longitudinal electric
field in question is entirely electrostatic.
The Hamiltonian, corresponding to the Lagrangian~(\ref{L1}) reads              
\begin{eqnarray}
H[{\bpi}, {\bA}; p,q] &=& \bp \cdot 
\dot{\bq} +\int {\bpi}\cdot \dot{\bA} d\br - L  \\
&=& \frac{1}{2m} 
\bigl[{\bp} - \frac{e}{c}{\bA}({\bq}, t) \bigr]^2 +
 \frac{1}{8\pi} \int \bigl[(4\pi c {\bpi})^2 
+ (\mbox{curl} {\bA})^2\bigr] d\br \nonumber
\end{eqnarray} 
where the corresponding conjugate momenta are  
${\bp} = m \dot{\bq} + (e/c) \bA (\bq, t), \quad
{\bpi}(\br) = (4\pi c^2)^{-1} \dot{\bA}.$

It is well known that in classical dynamics the addition of a total
time derivative to a Lagrangian leads to a new Lagrangian with the 
equations of motion unaltered. Lagrangians obtained in this manner are 
said to be equivalent. In general, the Hamiltonians following from the 
equivalent Lagrangians are different. Even the relationship between the 
conjugate and the kinetic momenta may be changed~\cite{PT}.
Let us now construct an equivalent Lagrangian to that of~(\ref{L1}) 
in the following way  
\begin{equation} 
L_1 = L - \frac{1}{c}\frac{d}{dt} \int \bP({\br})
\cdot{\bA}({\br}) d\br - \int {\rm div} (\bM \times \bA) d\br
\label{equiv}
\end{equation}
The new Lagrangian is a function of the variables $\bq$,
$\dot{\bq}$ and a functional of the field variables $\bA$, 
$\dot{\bA}$, and the equations of motion follow from the variational 
principle. We have
\begin{eqnarray} 
\frac{\p L_1}{\p \bA_i} &=& \frac{1}{c} \bigl(\bJ - \dot{\bP} -
c\,{\rm curl} \bM \bigr)_i = \bj_i \nonumber\\
\sum_{j}\frac{\p}{\p x_j}\frac{\p L_1}{\p (\p \bA_i/\p x_j)} &=& 
\frac{1}{4 \pi} \bigl[{\rm curl} \bB -  4 \pi {\rm curl} \bM\bigr]_i 
= \frac{1}{4 \pi} \bigl[{\rm curl} \bH\bigr]_i \nonumber \\
\frac{\p}{\p t}\frac{\p L_1}{\p (\p \bA_i/\p t)} &=&
-\frac{1}{4 \pi c} \frac{\p}{\p t} \bigl[ \bE + 4 \pi \bP\bigr]_i =
-\frac{1}{4 \pi c} \frac{\p \bD_i}{\p t} \nonumber
\end{eqnarray} 
Applying the Euler-Lagrange equations of motion 
\begin{eqnarray}
\frac{\p}{\p t}\,
\frac{\p L_1}{\p \dot{\bA}_i({\br})} + \sum_{j}
\frac{\p}{\p x_j}
\frac{\p L_1}{\p (\p
\bA_i/\p x_j)} -
\frac{\p L_1}{\p {\bA}_i} = 0 \nonumber
\end{eqnarray}
we obtain~(\ref{MMa}), i.e.,
\begin{eqnarray}
{\rm curl \bH} = \frac{1}{c} \frac{\p \bD}{\p t} + \frac{4\pi}{c} \bj \nonumber
\end{eqnarray}
The corresponding Hamiltonian now can be written as
\begin{eqnarray}
H_1[{\bpi}, {\bA}; p,q] &=& 
\frac{1}{2m} 
\bigl[{\bp} - \frac{e}{c}{\bA}({\bq}, t) \bigr]^2 
+  \frac{1}{8\pi} \int \bigl[\bD^2 + \bB^2\bigr] d\br \\
&-& \int \bD \cdot 
\bP d\br + 2 \pi \int \bP^2 d\br  - \int \bM \cdot \bB d\br \nonumber
\end{eqnarray} 
In the same way we can introduce the toroid polarizations in the system.
If we take into account that the total current now has the form~\cite{DSh}
\begin{equation}
\bj_{\rm total} = \bJ = \bj + \frac{\p \bP}{\p t} + 
\frac{\p {\rm curl} \bT^e}{\p t} + c\, {\rm curl} \bM + 
c\, {\rm curl\,curl} \bT^m \label{tc}
\end{equation}
we again obtain~(\ref{MMa}) with
$\bD = \bE + 4\pi (\bP + {\rm curl} \bT^e)$ and
$\bH = \bB - 4\pi (\bM + {\rm curl} \bT^m)$.
Here $\bT^e$ and $\bT^m$ are the toroid polarizations of electric and 
magnetic type respectively. In terms of $\bD$ and $\bB$ the system~
(\ref{Eq:MM}) can be written as follows
\begin{mathletters}
\label{New1}
\begin{eqnarray}
{\rm curl} \cB &=& \frac{1}{c}\frac{\p \cD}{\p t}  
+ 4 \pi\{{\rm curl}\,\bM + {\rm curl\,curl}\,\bT^m \} + \frac{4 \pi}{c} \bj
\label{Ne1}\\  
{\rm div} \cD &=& 4 \pi \rho \label{Ne2} \\
{\rm curl} \cD &=& - \frac{1}{c}\frac{\p \cB}{\p t} 
+ 4 \pi\{ {\rm curl}\,\bP + {\rm curl\,curl}\,\bT^e\} \label{Ne3}   \\
{\rm div} \cB &=& 0 \label{Ne4}  
\end{eqnarray}
\end{mathletters}
where we forced to redefine $\bB$ and $\bD$ as $\cB$ and $\cD$ respectively.
Indeed, due to the introduction of toroid polarizations, having independent
origin in terms of atomic and molecular current and charge destributions, 
the quantities $\bB$ and $\bD$ as well as $\bE$ and $\bH$ lost their initial 
meaning. The existence of the vorticities $\bT^e$ and $\bT^m$, generally
speaking, can be imputed to the one and the same physical volume. So in what 
follows, we substitute $\bE,\, \bB,\, \bD,\, \bH$ by $\cE,\,\cB,\,\cD,\,\cH$.

The equation of motion
(~\ref{LF}) should also be rewritten as follows
\begin{equation}
\label{NLF} 
m \ddot{\bq} = e {\cE} + \frac{e}{c}\dot{\bq} \times {\cB}
\end{equation}
The equation ~(\ref{Ne1}) can be derived as earlier by constracting a 
Lagrangian equivalent to $L_1$ such that
\begin{equation} 
L_2 = L_1 - \frac{1}{c}\frac{d}{dt} \int {\rm curl} \bT^e({\br})
\cdot{\bA}({\br}) d\br - \int {\rm div} ({\rm curl}\bT^m \times \bA) d\br
\label{equivalent}
\end{equation}
The corresponding Hamiltonian reads 
\begin{eqnarray}
H_2[{\bpi}, {\bA}; p,q] &=& 
\frac{1}{2m} 
\bigl[{\bp} - \frac{e}{c}{\bA}({\bq}, t) \bigr]^2 +
\frac{1}{8\pi} \int \bigl[\cD^2 + \cB^2\bigr] d\br  \\
&-&\int \cD \cdot [\bP  + {\rm curl} \bT^e] d\br 
+ 2 \pi \int [\bP + {\rm curl} \bT^e]^2 d\br  - 
\int [\bM + {\rm curl} \bT^m] \cdot \cB d\br  \nonumber
\end{eqnarray} 
Note that in our previous work we impose the following additional condition 
\begin{equation}
{\rm curl}\, \bT^{m,e} = \pm \frac{1}{c} 
\dot{\bT}^{e,m}
\label{AC}
\end{equation}
The relation~(\ref{AC}) demands some comments. Both $\bT^e$ and $\bT^m$ 
represent the closed isolated lines of electric and magnetic fields. So 
they have to obey the usual differential relations similar to the free 
Maxwell equations~\cite{DK,Miller}). However, 
remark that signs here are opposite to the corresponding one in
Maxwell equations because the direction of electric dipole 
is accepted to be chosen opposite to its inner electric field~\cite{Ginzburg}.
It should be emphasized that the relation~(\ref{AC}) is a local one and
it is not necessary to demand the condition~(\ref{AC}) to be held to 
introduce toroid polarizations to the electromagnetic equations.
We would also like to remark that the equations ~(\ref{Ne1}) and ~(\ref{Ne3})
can be derived strait from ~(\ref{MMa}) and ~(\ref{MMc}) making the following
substitutions in them:
\begin{eqnarray}
\bP \Longrightarrow \bP + {\rm curl} \bT^e, \quad
\bM \Longrightarrow \bM + {\rm curl} \bT^m \nonumber
\end{eqnarray} 
and redefining the vectors $\bB$ and $\bD$ as earlier.
\section{Two Potential Formalism}
\setcounter{equation}{0}
It is generally inferred that the divergence equations of the Maxwell
system are "redundant" since they are the consequences of curl equations
under the condition of coninuity~\cite{Stratton}. Recently Krivsky a.o.
~\cite{Krivsky} 
claimed that to describe the free electromagnetic field it is sufficient 
to consider the curl-subsystem of Maxwell equations since the equalities 
${\rm div} \bE = 0$ and ${\rm div} \bB = 0$ are fulfilled identically. 
Contrary to this statement, it has been proved that the 
divergence equations are not redundant and that neglecting these equations 
is at the origin of spurious solutions in computational electromagnetics
\cite{Jiang,Hillion}. 
Here we construct generalized formulation of Maxwell equations including 
both curl and divergence subsystems. In this section we develop two potential 
formalism (a similar formalism was developed by us earlier with the 
curl-subsystem taken into account only). Note that in the ordinary one 
potential formalism ($\bA, \vf$) the second set of Maxwell equations are 
fulfilled identically. So that all the four Maxwell equations bring their 
contribution individually, in our view, one has to rewrite the Maxwell 
equation in terms of two vector and two scalar potentials. To this end we 
introduce so-called double potential ~\cite{DMag,Kluwer,ICTP} 
to the system~(\ref{New1}), i.e.,
\begin{mathletters}
\label{New}
\begin{eqnarray}
{\rm curl} \cB &=& \frac{1}{c}\frac{\p \cD}{\p t} + \frac{4 \pi}{c} 
\bj_{\rm free} + 4 \pi\{ {\rm curl}\,\bM + {\rm curl\,curl}\,\bT^m \} 
\label{N1}\\  
{\rm div} \cD &=& 4 \pi \rho \label{N2} \\
{\rm curl} \cD &=& - \frac{1}{c}\frac{\p \cB}{\p t} 
+ 4 \pi\{ {\rm curl}\,\bP + {\rm curl\,curl}\,\bT^e\} \label{N3}   \\
{\rm div} \cB &=& 0 \label{N4}  
\end{eqnarray}
\end{mathletters}
Before developing the two potential formalism we first rewrite 
system~(\ref{Eq:M}) in terms of vector and scalar potentials $\bA, \phi$ 
such that 
$\bB = {\rm curl}\,\bA$, $\bE = -\nabla \phi - (1/c)(\p \bA/ \p t)$.
Following any text book we can write system~(\ref{Eq:M}) as 
\begin{mathletters}
\label{Eq:al7}
\begin{eqnarray}
\Box\, \bA &=& -\frac{4\pi}{c} \bj_{\rm tot} = -\frac{4\pi}{c}
\bigl[\bj_{\rm free} + \frac{\p \bP}{\p t} + c\, {\rm curl}\, \bM \bigr] 
\label{E7a}\\
\Box \phi &=& -4\pi \bigl[\rho -{\rm div}\,\bP \bigr] \label{E7b}
\end{eqnarray}
\end{mathletters}
under Lorentz gauge, i.e., 
${\rm div}\, \bA + (1/c) (\p \phi/\p t) = 0$ and
\begin{mathletters}
\label{Eq:al9}
\begin{eqnarray}
\Box \bA &=& -\frac{4\pi}{c} \bigl[\bj_{\rm tot} - \frac{1}{4 \pi} 
\nabla \frac{\p \phi}{\p t}\bigr] \label{E9a}\\
\nabla^2 \phi &=& -4\pi \bigl[\rho -{\rm div}\bP \bigr] \label{E9b}
\end{eqnarray}
\end{mathletters}
under Coulomb gauge, i.e., ${\rm div} \bA = 0$.
Here $\Box = \nabla^2 - (1/c^2) (\p^2 /\p t^2).$ 
Note that to obtain~(\ref{Eq:al7}) or~(\ref{Eq:al9}) it is sufficient to 
consider~(\ref{Ma}) and~(\ref{Mb}) only since the two others are fulfilled 
identically. 

Let us now develop two potential formalism. Two potential formalism was 
first introduced in~\cite{DMag} and further developed in~\cite{Kluwer,ICTP}. 
In both papers we introduce only two vector potentials 
$\bal^m,\,\bal^e$ and use only the curl-subsystem of the Maxwell equations 
with the additional condition ${\rm  div} \bal^{m, e} = 0$. 
Thus, in our view our previous version of two potential formalism lack of 
completeness. In the present paper together with the vector potentials 
$\bal^m, \bal^e$ we introduce two scalar potentials $\vf^m$ and 
$\vf^e$ such that 
\begin{mathletters}
\begin{eqnarray}
\cB &=&
{\rm curl}\bal^m + \frac{1}{c} \frac{\p \bal^e}{\p t} + \nabla \vf^m, \\
\cD &=&
\mbox{curl}\bal^e - \frac{1}{c}\frac{\p \bal^m}{\p t} - \nabla \vf^e 
\end{eqnarray}
\end{mathletters}
It can be easily verified that system of equations~(\ref{New}) 
are invariant under this transformation and take the form 
\begin{mathletters}
\label{LG}
\begin{eqnarray}
\Box\, \bal^m &=& - \frac{4 \pi}{c}\bigl[\bj + c\, {\rm curl}\, \bM + 
c\, {\rm curl\,curl}\,\bT^m\bigr], \\ 
\Box\, \vf^m &=& 0 \\
\Box\, \bal^e &=& -\frac{4 \pi}{c}\bigl[{\rm curl}\, \bP +  
{\rm curl\,curl}\,\bT^e\bigr], \\ 
\Box\, \vf^e &=& - 4 \pi\, \rho 
\end{eqnarray}
\end{mathletters}
under 
${\rm div}\,\bal^{m,e} + (1/c) (\p \vf^{e,m}/\p t) = 0$
and 
\begin{mathletters}
\label{CG}
\begin{eqnarray}
\Box\, \bal^m &=& - \frac{4 \pi}{c}\bigl[\bj + c\, {\rm curl}\, \bM + 
c\, {\rm curl\,curl}\,\bT^m - \frac{1}{4\pi} \nabla 
\frac{\p \vf^e}{\p t}\bigr] \label{E14a}\\ 
\nabla^2\, \vf^m &=& 0 \label{E14b}\\
\Box\, \bal^e &=& -\frac{4 \pi}{c}\bigl[{\rm curl}\, \bP +  
{\rm curl\,curl}\,\bT^e -\frac{1}{4\pi}\frac{\p \vf^m}{\p t} \bigr]
\label{E14c}\\ 
\nabla^2\, \vf^e &=& - 4 \pi\, \rho \label{E14d}
\end{eqnarray}
\end{mathletters}
under ${\rm div}\, \bal^{m,e} = 0.$
The solutions to the systems~(\ref{LG}) and~(\ref{CG}) can be written as
follows (see for example~\cite{ICTP,ODJ}):
The solutions to the d'Alembert equation
\begin{equation}
\Box F (\br, t)= f(\br, t)
\label{Alem}
\end{equation}
look
\begin{equation}                  
F (\br, t) = -\frac{1}{4\pi} \int\limits_{\mbox{all space}} \frac{f(\br', t') 
d\br'}{|\br - \br'|}\Biggr|_{t' = t - |\br - \br'|/ c} 
\label{}
\end{equation}
whereas the solutions to the Poisson equation 
\begin{equation}
\nabla^2 F(\br) = f(\br) 
\label{Pois}
\end{equation}
read
\begin{equation}
F (\br) = -\frac{1}{4\pi} \int \frac{f(\br') d\br'}
{|\br - \br'|}
\label{}
\end{equation}

Let us rewrite the equations~(\ref{LG}) and ~(\ref{CG}) for the fields
subject to Lorentz transformation. If the fields in stationary 
frame (unprimed) are connected with those in moving one (primed) in the
following way
\begin{mathletters}
\label{moving}
\begin{eqnarray}
\bB &=& \gamma\bigl(\bB^{\pr} + \frac{1}{c} \bbe \times \bE^{\pr} \bigr) \\
\bE &=& \gamma\bigl(\bE^{\pr} - c \bbe \times \bB^{\pr} \bigr) \\
\bP &=& \gamma\bigl(\bP^{\pr} + \frac{1}{c} \bbe \times \bM^{\pr} \bigr) \\ 
\bM &=& \gamma\bigl(\bM^{\pr} - c \bbe \times \bP^{\pr} \bigr) \\
\bT^e &=& \gamma\bigl(\bT^{e \pr} + \frac{1}{c}\bbe \times \bT^{m \pr}\bigr) \\
\bT^m &=& \gamma\bigl(\bT^{m \pr} - c \bbe \times \bT^{e \pr} \bigr) \\
\bal^m &=& \gamma\bigl(\bal^{m \pr} + \bbe \vf^{e \pr} \bigr) \\
\vf^m &=& \gamma\bigl(\vf^{m \pr} - \bbe \bal^{e \pr} \bigr) \\
\bal^e &=& \gamma\bigl(\bal^{e \pr} - \bbe \vf^{m \pr} \bigr) \\
\vf^e &=& \gamma\bigl(\vf^{e \pr} + \bbe \bal^{m \pr} \bigr) \\
\rho &=& \gamma\bigl(\rho^{\pr} + \frac{1}{c} \bbe \bJ^{\pr} \bigr) \\ 
\bJ &=& \gamma\bigl(\bJ^{\pr} + c \bbe \rho^{\pr} \bigr) 
\end{eqnarray} 
\end{mathletters}
The equations ~(\ref{LG}) now can be written as
\begin{mathletters}
\label{LG1}
\begin{eqnarray}
\Box\, \bal^{m \pr}&=& - \frac{4 \pi \gamma^2}{c}\bigl[\gamma^{-1}\bj^{\pr} 
+ c\, {\rm curl}\, \bigl(\bM^{\pr} + {\rm curl}\,\bT^{m \pr}\bigr)
- {\rm curl}\, \bigl(\bbe \times \bP^{\pr} + {\rm curl}\,(\bbe\times
\bT^{e \pr})\bigr) \bigr], \\ 
\Box\, \vf^{m \pr}&=& \bbe \Box \bal^{e \pr} \\
\Box\, \bal^{e \pr}&=& - \frac{4 \pi \gamma^2}{c}\bigl[ 
{\rm curl}\, \bigl(\bP^{\pr} + {\rm curl}\,\bT^{e \pr}\bigr)
+\frac{1}{c} {\rm curl}\, \bigl(\bbe \times \bM^{\pr} + 
{\rm curl}\,(\bbe\times \bT^{m \pr})\bigr) \bigr], \\ 
\Box\, \vf^{e \pr}&=& -\bbe \Box \bal^{m \pr} - 4 \pi (\rho^{\pr} +
\bbe \bJ^{\pr}/c)
\end{eqnarray}
\end{mathletters}
It is necessary to emphasize that the potential descriptions 
electrotoroidic and magnetotoroidic media are completely separated. The 
properties of the magnetic and electric potentials $\bal^m$ and $\bal^e$ 
under the temporal and spatial inversions are opposite \cite{DK}. 
The potential $\bal^e$ ($\bal^m$) is related to the toroidness of the
medium $\bT^e$ ($\bT^m$) as $\bB$ ($\bD$) to $\bM$ ($\bP$). 
\section{Conclusion}
\setcounter{equation}{0}
The modified equations of electrodynamics has been obtained in account
of toroid moment contributions. The two-potential formalism has been
further developed for the equations obtained. Note that introduction
of free magnetic current $\bj_{\rm free}^{m}$ and magnetic charge
$\rho^{m}$ in the equations~(\ref{N3}) and ~(\ref{N4}) respectively
leads to the equations obtained by Singleton~\cite{Singleton1}.

\end{document}